\documentclass[a4paper,12pt]{article}
\usepackage{amsthm}
\usepackage{amsmath}
\usepackage{amssymb}
\usepackage{color}
\usepackage{graphicx}
\usepackage{subfigure}
\usepackage{cite}
\usepackage[linktocpage]{hyperref}
\usepackage{setspace}

\newcommand{\mrd}{\mathrm{d}}

\begin{document}
\let\endtitlepage\relax

\begin{titlepage}
\begin{center}

\newcommand\blfootnote[1]{%
	\begingroup
	\renewcommand\thefootnote{}\footnote{#1}%
	\addtocounter{footnote}{-1}%
	\endgroup
}

\vspace*{-1.0cm}
\renewcommand{\baselinestretch}{1.0}  
\setstretch{1.6}
{\Large{\textbf{Hodge Dual Gauge Symmetry in Minimal Einstein-Aether Theory}}}

\renewcommand{\baselinestretch}{1.0}  
\setstretch{1.2}

\vspace{8mm}
\centerline{\large{Kamal Hajian\blfootnote{kamal.hajian@uni-oldenburg.de} }}
\vspace{-1mm}
\normalsize
\textit{Department of Physics, Middle East Technical University, 06800, Ankara, Turkey}\\
\textit{Institute of Physics, University of Oldenburg, P.O.Box 2503, D-26111 Oldenburg, Germany}

\renewcommand{\baselinestretch}{1.0}  
\setstretch{1}

\begin{abstract}
Einstein-aether gravity is a theory that breaks the local Lorentz symmetry by introducing a preferred direction via a vector field, which is considered to play the role of an aether. The theory is identified by four coupling constants between the aether and gravity. Minimal Einstein-aether is the special case in which only one of the couplings is non-zero. We show that the aether vector field in its minimal version is Hodge dual to a gauge field. The gauge symmetry in the dual description has been known for decades and has been used to implement a cosmological constant into the Lagrangian. As a result, solutions to the well-established gauge theory can be transferred into the minimal Einstein-aether theory straightforwardly. On the other hand, some of the proposed solutions to the minimal Einstein-aether theory could be discarded as pure gauges of the vanishing aether. We prove as a theorem that this holds true for all divergence-less aether fields. 
\end{abstract}
\end{center}
\vspace*{0cm}
\end{titlepage}
\vspace*{-0mm}

\section{Introduction}
Einstein-aether theory is a modification of general relativity that introduces a dynamical (unit timelike) vector field -called the aether- into spacetime \cite{Jacobson:2000xp,Jacobson:2007veq,Jacobson:2013xta}. This vector field defines a preferred local rest frame at every point in space and time, effectively breaking local Lorentz invariance \cite{Mattingly:2005re,Liberati:2013xla}, a key symmetry in Einstein's theory of relativity. The theory was developed to explore possible violations of Lorentz symmetry, which some quantum gravity and high-energy physics models suggest might occur at very small scales or high energies. Despite this breaking of Lorentz invariance, Einstein-aether theory is still generally covariant, meaning it respects coordinate invariance like general relativity.

In terms of physics, the aether field interacts with the spacetime metric and contributes to the gravitational dynamics. The theory introduces several new parameters that govern the behavior of this field, and constraints on these parameters come from observations like gravitational waves, cosmology, and the behavior of black holes and neutron stars (see Refs. \cite{Jacobson:2007fh,Adam:2021vsk,Gupta:2021vdj,Oost:2018tcv,Foster:2005dk,Elliott:2005va} and references therein). Recent decades have seen investigations into other facets of the theory, including black holes  \cite{Eling:2006ec,Garfinkle:2007bk,Tamaki:2007kz,Barausse:2011pu,Gao:2013im,Ding:2015kba,Adam:2021vsk,Zhang:2020too} and other types of solutions \cite{Gurses:2015bjt}, stability \cite{Donnelly:2010qd,Carroll:2009em,Tsujikawa:2021typ}, cosmological implications \cite{Carroll:2004ai,Zlosnik:2007bu,Zuntz:2010jp,Zlosnik:2006zu}, generalizations \cite{Bonvin:2007ap,Balakin:2014tza,Alpin:2015nen}, and spacelike/null aether fields \cite{Carroll:2008pk,Ackerman:2007nb,Chatrabhuti:2009ew,Gurses:2016vwm}.

In a distinct and significantly earlier line of research, Einstein introduced the cosmological constant $\Lambda$ \cite{Einstein:1917} within general relativity to explain the seemingly static universe. However, with the subsequent discovery of the Universe's accelerating expansion \cite{Perlmutter:1998np,Riess:1998cb} and the advent of the AdS/CFT correspondence \cite{Maldacena:1997re,Brown:1986nw}, $\Lambda$ has become a central focus in cosmology. In Einstein's original formulation, the cosmological constant was incorporated as a fundamental, constant parameter within the Lagrangian; essentially part of the theory's definition. A less conventional approach \cite{Aurilia:1980xj, Duff:1980qv} allows $\Lambda$ to emerge as a free parameter in the solution by introducing a new gauge field into the Lagrangian. The formulation was further developed in the early 1980s through thorough investigations of its Hamiltonian dynamics, which identified the canonical variables (the field and its conjugate momentum) and constants of integration \cite{Henneaux:1984ji,Henneaux:1985tv,Teitelboim:1985dp,Henneaux:1989zc}.  This line of investigation was further advanced by demonstrating that $\Lambda$ is not only a constant of integration in the solution but also a conserved charge linked to the global component of the gauge symmetry of this new gauge field \cite{1,Hajian:2021hje}. More recently, this gauge field formulation has been extended to allow not only the cosmological constant but also other coupling constants to be promoted to integration constants and conserved charges \cite{Hajian:2023bhq}.

Here, the aim is to demonstrate that the Einstein-aether gravity and the gauge field formulation of the cosmological constant, two seemingly distinct lines of research, intersect at a specific point: the ``minimal Einstein-aether" theory. More precisely, we will show that these two frameworks are dual to each other via a Hodge transformation. To achieve this, the following sections will proceed as follows: Sec. \ref{reviews} will provide a brief review of both models; Section \ref{duality} will introduce the duality between them; Section \ref{implications} will explore some applications of the duality; and finally, Section \ref{conclusion} will summarize our analysis.

\noindent\underline{Conventions:} We focus on $4$-dimensional spacetime though the generalization to arbitrary dimensions is straightforward. The signature of the metric $g_{\mu\nu}$ is chosen to be $(-,+,+,+)$ in this paper, and we set the light speed and Planck constant to unity, $c=\hbar=1$. The Levi-Civita tensor is the volume form denoted by $\epsilon_{\mu_{_1}\mu_{_2}\mu_{_3}\mu_{_4}}$ with the sign convention $\epsilon_{012 3}=+\sqrt{-g}$. The notation $[\mu_{_1}\mu_{_2}\dots\mu_p]$ will be used to denote antisymmetrization over the set of indices within the bracket normalized by the factor $\frac{1}{p!}$. The exterior derivative of a $p$-form ${a}=\frac{1}{p!}a_{\mu{_1}\dots \mu{_p}}dx^{\mu_{_1}}\wedge\dots\wedge dx^{\mu_p}$ (for $0\leq p<4$) is defined as
\begin{equation}
\mrd{a} \equiv (p+1)\,\nabla_{[\mu}a_{\mu_{_1}\dots\mu_{p}]}\,\mrd x^{\mu}\wedge\mrd x^{\mu_{_1}}\wedge\dots\wedge \mrd x^{\mu_{p}},
\end{equation}
which in particular for $p = 4$ implies the identity $\mrd a = 0$. Hodge duality of ${a}$ (for $0<p<4$), which is denoted as $\star{a}$, is a $(4\!-\!p)$-form  defined as 
\begin{align}\label{Hodge duality}
\left(\star{a}\right)_{\mu_{p+1}\dots\mu_{_4}}=\frac{1}{p!}\epsilon_{\mu_{_1}\dots\mu_p\mu_{p+1}\dots\mu_{_4}}\,a^{\mu_{_1}\dots\mu_p}\,,
\end{align}
which for $p = 0$ and $p = 4$ becomes
\begin{align}\label{Hodge scalar}
\left(\star{a}\right)_{\mu_{_1}\mu_{_2}\mu_{_3}\mu_{_4}}=\epsilon_{\mu_{_1}\mu_{_2}\mu_{_3}\mu_{_4}}a, \qquad \left(\star{a}\right)=\frac{1}{4!}\epsilon_{\mu_{_1}\mu_{_2}\mu_{_3}\mu_{_4}}\,a^{\mu_{_1}\mu_{_2}\mu_{_3}\mu_{_4}}.
\end{align}
In the equations that contain $\pm$ and $\mp$, the upper signs will be understood to be in correspondence with each other, and similarly for the lower signs.

\section{Quick review of the two theories}\label{reviews}
\subsection{Review 1: Minimal Einstein-aether theory}
A generic aether theory in four dimensions can be described by the generally covariant action
\begin{equation}\label{aether action}
I=\frac{1}{16\pi G}\int \mrd^4 x \sqrt{-g} \left( \mathcal{L}_g+ \mathcal{L}_M-K^{\mu\nu}_{\;\;\,\;\alpha\beta} \nabla_\mu u^\alpha \nabla_\nu u^\beta +\lambda (u_\sigma u^\sigma+1)\right),
\end{equation}
in which $\mathcal{L}_g$ and $\mathcal{L}_m$ denote the gravity and matter Lagrangian densities respectively, and
\begin{equation}
K^{\mu\nu}_{\;\;\,\;\alpha\beta}=c_1 g^{\mu\nu}g_{\alpha\beta}+c_2 \delta^\mu_\alpha \ \delta^\nu_\beta+c_3 \delta^\nu_\alpha \ \delta^\mu_\beta-c_4 u^\mu u^\nu g_{\alpha\beta}.
\end{equation}
The coefficients $\{c_1,c_2,c_3, c_4\}$ are dimensionless constants in the theory which couple the aether field $u^\mu$ to the metric $g_{\mu\nu}$ such that the field equations are at most second order. By the effective field theory considerations \cite{Withers:2009qg} the $\mathcal{L}_g$ is determined to be the curvature scalar $R$, yielding to the Einstein-aether theory. The $\lambda$ in the last term of the action \eqref{aether action} is a Lagrange multiplier that constrains the aether to be everywhere non-vanishing and time-like. The special case of $c_1=c_3=c_4=0$ and $c_2\neq 0$ is called ``reduced Einstein-aether" theory \cite{Gurses:2024jkx}, whose Lagrangian is
\begin{equation}\label{reduced}
\mathcal{L}=R+\mathcal{L}_M-c_2(\nabla_\alpha u^\alpha)^2+\lambda (u_\sigma u^\sigma+1),
\end{equation} 
which is favored by the observational constraints (see Ref. \cite{Gurses:2024jkx} and references therein). By the principle of least action, the field equations of the reduced Einstein-aether model are  
\begin{align}
&G_{\mu\nu}=8\pi G\ T_{\mu\nu}+\left( \lambda +\frac{c_2}{2}(\nabla_\alpha u^\alpha)^2\right) g_{\mu\nu}+\lambda u_{\mu}u_{\nu}, \label{aether EoM g}\\
&\nabla_\mu (\nabla_\alpha u^\alpha)+\frac{\lambda}{c_2}u_\mu=0 \qquad \qquad (c_2\neq 0), \label{aether EoM u}
\end{align}
in which $G_{\mu\nu}$ is the Einstein tensor, and $T_{\mu\nu}$ is the  matter energy-momentum tensor, 
\begin{equation}
T_{\mu\nu}\equiv\frac{-1}{8\pi G\sqrt{-g}} \frac{\delta (\sqrt{-g}\mathcal{L}_M)}{\delta{g^{\mu\nu}}}.
\end{equation}
Contracting both sides of Eq. \eqref{aether EoM u} by $u^\mu$ and using $u_\sigma u^\sigma=-1$, it can be rewritten as
\begin{equation}\label{lambda}
\lambda =c_2 u^\mu\nabla_\mu (\nabla_\alpha u^\alpha).
\end{equation}
Notice that in the derivation of Eq. \eqref{aether EoM g}, the relation \eqref{lambda} has been used. 
 
The ``minimal Einstein-aether" theory is achieved from the reduced theory by requesting additional conditions such as $\nabla_\alpha u^\alpha=\text{const.}$ or $\lambda=0$. However, to prevent any confusion, we crystallize different possibilities to implement such conditions. 
\begin{itemize}
\item[(I)] $\boldsymbol{\nabla_\alpha u^\alpha=}\textbf{const.}$ \textbf{off-shell:} In this case, the term $c_2(\nabla_\alpha u^\alpha)^2$ in the reduced Lagrangian Eq.\eqref{reduced} can be replaced by a constant $2\tilde \Lambda$ which results to
\begin{equation}\label{L I}
\mathcal{L}=R+\mathcal{L}_M-2\tilde \Lambda+\lambda (u_\sigma u^\sigma+1),
\end{equation}
via $\tilde \Lambda=\frac{c_2(\nabla_\alpha u^\alpha)^2}{2}$. This Lagrangian is simply the general relativity with a cosmological constant $\tilde\Lambda$ and the matter field without any dynamical and interacting aether. The vector $u^\mu$ does not enter into the field equations, and it is a redundancy in the theory which makes the condition $u_\sigma u^\sigma=-1$ an irrelevant constraint. So, the case (I) may not be considered as an Einstein-aether theory. 

\item[(II)] $\boldsymbol{\lambda=0}$ \textbf{off-shell:} The achieved theory is described by
\begin{equation}\label{II L}
\mathcal{L}=R+\mathcal{L}_M-c_2(\nabla_\alpha u^\alpha)^2,
\end{equation}
which is the reduced Einstein-aether theory with an aether field $u^\mu$ that is no longer constrained to be timelike and non-vanishing everywhere. The field equations in this theory are the same as the Eqs. \eqref{aether EoM g} and \eqref{aether EoM u} with $\lambda=0$, i.e.,
\begin{align}
&G_{\mu\nu}=8\pi G\ T_{\mu\nu}+\frac{c_2}{2}(\nabla_\alpha u^\alpha)^2 g_{\mu\nu}, \label{II EoM g}\\
&\nabla_\mu (\nabla_\alpha u^\alpha)=0. \label{II EoM u}
\end{align}
Equation \eqref{II EoM u} implies $\nabla_\alpha u^\alpha=\text{const.}$ on-shell. So, the term $\frac{c_2}{2}(\nabla_\alpha u^\alpha)^2$ in \eqref{II EoM g} can be replaced by a constant via 
\begin{equation}\label{tilde Lambda}
{\Lambda}\equiv-\frac{c_2}{2}(\nabla_\alpha u^\alpha)^2,
\end{equation}
which simplifies \eqref{aether EoM g} to
\begin{align}\label{EoM Lambda tilde}
&G_{\mu\nu}+ \Lambda g_{\mu\nu}=8\pi G\ T_{\mu\nu}.
\end{align} 
This equation shows that the case (II) reproduces the field equations of the  Einstein-$\Lambda$ theory on-shell.  

\item[(III)] $\boldsymbol{\nabla_\alpha u^\alpha=}\textbf{const.}$ \textbf{or} $\boldsymbol{\lambda=0}$ \textbf{on-shell:} The Lagrangian of the theory is the same as the reduced Einstein-aether in Eq. \eqref{reduced}. In this respect, it may not be appropriate to distinguish the case (III) as a different model or theory (under the name of e.g., minimal Einstein-aether \emph{theory}).  However, an extra condition, either ${\nabla_\alpha u^\alpha=}{\text{const.}}$ or $\lambda=0$, is aimed to be imposed on the solutions. To this end, one of the conditions is inserted in the field equation  \eqref{aether EoM u}.  It is clear from Eq. \eqref{aether EoM u} that requesting any one of these two conditions yields the other one. Introducing the constant $\Lambda$ via \eqref{tilde Lambda}, and by replacing the ${\nabla_\alpha u^\alpha=}{\text{const.}}$ and $\lambda=0$ in Eq. \eqref{aether EoM g} an equation identical to Eq. \eqref{EoM Lambda tilde} is achieved. Therefore, the solutions in the case (III) satisfy the same field equations \eqref{tilde Lambda} and \eqref{EoM Lambda tilde} but with the constraint $u_\alpha u^\alpha=-1$.  
\end{itemize}

Some comments are helpful in the clarification of the above discussion. 
\begin{itemize}
\item Any solution in (III) is also a solution to the case (II). To see this, we note that the field equations in both scenarios are similar, that are Eqs. \eqref{tilde Lambda} and \eqref{EoM Lambda tilde}. The only difference is that the aether field in (II) is not constrained to be timelike and non-vanishing.
\item In Ref. \cite{Gurses:2024jkx} the case (III) is identified as the ``minimal Einstein-aether" theory, while in Ref. \cite{Franzin:2023rdl} the same case that is delimited to $\nabla_\alpha u^\alpha=0$ is called minimal. 
\item We did not include any cosmological constant $\Lambda_0$ directly in the Lagrangian via 
\begin{equation}
\mathcal{L}\to \mathcal{L}-2\Lambda_0
\end{equation}
from the beginning. However, it could be added to the Einstein-aether generic Lagrangian \eqref{aether action}, and the analysis would follow straightforwardly with the replacement $\Lambda\to \Lambda_0+\Lambda$ in the field equations. It is important to note that in either (II) or (III), observational constraints delimit $c_2$ through the combination $\Lambda_0+ \Lambda$. Importantly, in this paper we have assumed $\Lambda_0=0$, and so, $c_2$ can take negative values (in an expanding universe where $\Lambda_0+ \Lambda\geq 0$). 
\item One can scale $u^\mu \to \sqrt{|c_2|}{u^\mu}$ in relations \eqref{II L}, \eqref{II EoM g}, and \eqref{II EoM u}. As a result of such a field redefinition, the coupling $c_2$ in those equations can  be set to $+1$ or $-1$ depending on the sign of $c_2$.
\end{itemize}
Our focus in this paper will be on the case (II), whose special subsets are called minimal theory in Refs. \cite{Gurses:2024jkx,Franzin:2023rdl}.

\subsection{Review 2: Gauge field formulation of cosmological constant}
In order to make the connection with the analysis in the previous section, we review the gauge formulation of the cosmological constant for the Einstein gravity and in four dimensions, although the formulation works in any dimensions and for any theory of gravity (reviewed in \cite{Hajian:2021hje}). Considering Einstein Lagrangian density $\mathcal{L}_g=R$ without a cosmological constant in $4$-dimensional spacetime with the matter Lagrangian density $\mathcal{L}_M$, the action and the gravitational sector of the field equations are 
\begin{equation}
I=\frac{1}{16\pi G}\int \mrd^4 x \sqrt{-g}\mathcal{L}=\frac{1}{16\pi G}\int \mrd^4 x \sqrt{-g}\left(R+\mathcal{L}_M\right), \qquad  G_{\mu\nu}=8\pi G\ T_{\mu\nu}.    
\end{equation}
The cosmological constant can be added to this model by introducing a gauge field $A_{\mu_{_1} \mu_{_2} \mu_{_3}}$ and its field strength $F_{\mu_{_1}\mu_{_2}\mu_{_3}\mu_{_4}}$, with the relation $F=\mrd A$, i.e., 
\begin{align}
F_{\mu_{_1}\mu_{_2}\mu_{_3}\mu_{_4}}&=4\nabla_{[\mu_{_1}} A_{\mu_{_2} \mu_{_3} \mu_{_4}]}\label{F dA}\\
&\equiv\frac{1}{3!}\left(\nabla_{\mu_{_1}} A_{\mu_{_2} \mu_{_3} \mu_{_4}}-\nabla_{\mu_{_2}} A_{\mu_{_1} \mu_{_3} \mu_{_4}}+\nabla_{\mu_{_2}} A_{\mu_{_3} \mu_{_1} \mu_{_4}}-\dots+\nabla_{\mu_{_4}} A_{\mu_{_3} \mu_{_2} \mu_{_1}}\right),  
\end{align}	
where ellipsis represents all other even and odd permutations of indices with plus and minus signs respectively. To this end, a new kinetic term is added to the Lagrangian 
\begin{equation}\label{L F2}
\mathcal{L}\to \mathcal{L}\mp 2 F^2 \quad \Rightarrow \quad  I \to I=\frac{1}{16\pi G}\int \mrd^4 x \sqrt{-g}\Big(R+\mathcal{L}_M  \mp 2 F^2 \Big),
\end{equation}
where $F^2 \equiv \frac{1}{4!}F_{\mu_{_1}\dots\mu_{_4}}F^{\mu_{_1}\dots\mu_{_4}}$. The field strength $F$ is a top-form, i.e., the number of its antisymmetric indices is equal to the spacetime dimension. We will see later that the $-$ and $+$ sign in front of the new term in the Lagrangian lead to a positive and negative cosmological constants respectively.  

In general, the field strength  $F$ can be a scalar function times the volume form, i.e. $F_{\mu_{_1}\dots\mu_{_4}}=\phi(x^\mu)\epsilon_{\mu_{_1}\dots\mu_{_4}}$. With variation of the action \eqref{L F2} with respect to $g_{\mu\nu}$ and $A$, the new field equations can be found as:
\begin{align}
&G_{\mu\nu}=8\pi G\ T_{\mu\nu}\pm \frac{1}{3} \Big(F_{\mu{\mu_{_2} \mu_{_3} \mu_{_4}}}F_{\nu}{}^{_{\mu_{_2} \mu_{_3} \mu_{_4}}}-3 F^2 g_{\mu\nu}\Big),\label{EoM gauge g}\\
&\nabla_\mu F^{\mu{\mu_{_2} \mu_{_3} \mu_{_4}}}=0. \label{EoM gauge F}
\end{align}
The most generic solution to the latter equation is $\phi(x^\mu)=\text{const.}\equiv c$ (see, e.g., Ref. \cite{Hajian:2021hje} to find the reason), and so,
\begin{equation}\label{F on-shell}
F_{\mu_{_1}\dots\mu_{_4}}=c\,\epsilon_{\mu_{_1}\dots\mu_{_4}} .   
\end{equation}
The constant $c$ can be assumed to be positive for convenience (not to be confused with the speed of light, which is set to 1). If this solution  is inserted in the other field equation \eqref{EoM gauge g}, it induces a cosmological constant:
\begin{equation}\label{EoM Lambda}
G_{\mu\nu}+\Lambda g_{\mu\nu}=8\pi G\ T_{\mu\nu},
\end{equation}
where
\begin{equation}\label{Lambda c}
\qquad\qquad  \Lambda =\pm c^2.
\end{equation}
Notice that in the derivation of the relations above the identities
\begin{equation}\label{identities}
\epsilon_{\mu{\mu_{_2}\mu_{_3}\mu_{_4}}}\epsilon_{\nu}{}^{{\mu_{_2}\mu_{_3}\mu_{_4}}}=-6g_{\mu\nu} \qquad \text{and}\qquad  \epsilon_{{\mu_{_1}\mu_{_2}\mu_{_3}\mu_{_4}}}\epsilon^{{\mu_{_1}\mu_{_2}\mu_{_3}\mu_{_4}}}=-24
\end{equation}
have been used. 

The improved action \eqref{L F2} has a gauge symmetry $A\to A+\mrd \alpha$ for a 2-form function $\alpha_{\mu\nu}$, i.e., $A_{\mu_{_1} \mu_{_2} \mu_{_3}}$ is modified to
\begin{equation}
A_{\mu_{_1} \mu_{_2} \mu_{_3}}\!+\frac{1}{2!}\left(\nabla_{\mu_{_1}}\alpha_{\mu_{_2} \mu_{_3}}\!-\nabla_{\mu_{_2}}\alpha_{\mu_{_1} \mu_{_3}}\!+\nabla_{\mu_{_2}}\alpha_{\mu_{_3} \mu_{_1}}\!-\nabla_{\mu_{_1}}\alpha_{\mu_{_3} \mu_{_2}}\!+\nabla_{\mu_{_3}}\alpha_{\mu_{_1} \mu_{_2}}\!-\nabla_{\mu_{_3}}\alpha_{\mu_{_2} \mu_{_1}}\right).
\end{equation}
The global part of this symmetry satisfies $\mrd \alpha=0$, i.e., it is the transformation which keeps $A$ intact.   By the Noether theorem and covariant formulation of charges \cite{Lee:1990gr,Wald:1993nt,Iyer:1994ys,Wald:1999wa,Ashtekar:1987hia,Ashtekar:1990gc,Crnkovic:1987at} (see Refs. \cite{Hajian:2015xlp,Hajian:2015eha} for reviews) one finds its associated conserved charge $C$ \cite{1,Hajian:2021hje} which is equal to
\begin{align}\label{C}
C= \pm \frac{\sqrt{|\Lambda|}}{4\pi G}  
\end{align}
in accordance with the $\mp$ signs in the improved Lagrangian \eqref{L F2}. Solutions to this theory, specifically black hole solutions and their conserved charges, have been investigated in details in the literature (see e.g., Ref. \cite{Hajian:2021hje}). 

Although replacing the solution Eq. \eqref{F on-shell} in the field equations yields the cosmological constant term, such a replacement in the Lagrangian \eqref{L F2} gives a $\Lambda$-like term with an incorrect sign. To remedy this issue, a surface term can be added to the Lagrangian as  \cite{Aurilia:1980xj,Duncan:1989ug,Wu}
\begin{align}\label{L F2 shifted}
I=\frac{1}{16 \pi G} \int \mrd ^4 x \sqrt{-g} \Big(R+\mathcal{L}_M  \mp 2 F^2\pm\frac{2 \nabla_\mu \left(A_{\mu_{_2} \mu_{_3} \mu_{_4}} F^{\mu{\mu_{_2} \mu_{_3} \mu_{_4}}}\right)}{3}\Big).
\end{align}
Applying Eq. \eqref{F dA}, Eq. \eqref{EoM gauge F}, and eventually Eq. \eqref{identities}, it is straightforward to check that this Lagrangian reduces to the standard Einstein-$\Lambda$ theory with matter field, i.e.,
\begin{equation}\label{L Einstein-Lambda}
\mathcal{L}=R+\mathcal{L}_M-2\Lambda, \qquad \qquad\qquad  \Lambda =\pm c^2
\end{equation}
if calculated on-shell via Eq. \eqref{F on-shell}.

\section{Equivalence of the two models}\label{duality}

In this section, we show that the case (II) of the minimal Einstein-aether theory is equivalent to the gauge formulation of the cosmological constant. To this end, we identify the 3-form field $A_{\mu_{_1} \mu_{_2} \mu_{_3}}$  to be the Hodge dual of the 1-form $u_\mu$ , i.e.,
\begin{equation}\label{identification}
A={\star \frac{u}{\sqrt{2}}} \quad \Rightarrow \qquad A_{\mu_{_1} \mu_{_2} \mu_{_3}}=\epsilon_{\mu\mu_{_1} \mu_{_2} \mu_{_3}}\frac{u^{\mu}}{\sqrt{2}},
\end{equation} 
where the norm of the aether field $u^\mu$ is fixed such that it makes $c_2=\mp 1$ in the Lagrangian \eqref{II L}. As a result of the identification \eqref{identification}, the Lagrangian \eqref{II L} in the minimal Einstein-aether theory is equal to the Lagrangian \eqref{L F2} via the off-shell relation
\begin{align}
&F^2=-\frac{(\nabla_\alpha u^\alpha)^2}{2}. \label{F2}
\end{align}
Besides, the Einstein-aether field equations \eqref{II EoM g} and \eqref{II EoM u} are equal to their analogues in the gauge formulation, i.e.,  Eqs. \eqref{EoM gauge g} and \eqref{EoM gauge F} using Eq. \eqref{F2} and the off-shell relation
\begin{align}
&F_{\mu{\mu_{_2} \mu_{_3} \mu_{_4}}}F_{\nu}{}^{_{\mu_{_2} \mu_{_3} \mu_{_4}}}=-3(\nabla_\alpha u^\alpha)^2 g_{\mu\nu}. \label{nabla A F}
\end{align}
To derive Eq. \eqref{F2}, we begin with the l.h.s and find the r.h.s as follows.
\begin{align}
F^2&=\frac{1}{4!}F_{\mu_{_1}\mu_{_2} \mu_{_3} \mu_{_4}}F^{\mu_{_1}\mu_{_2} \mu_{_3} \mu_{_4}}=\frac{1}{4!}(\mrd A)_{\mu_{_1}\mu_{_2} \mu_{_3} \mu_{_4}}(\mrd A)^{\mu_{_1}\mu_{_2} \mu_{_3} \mu_{_4}}\\
\eqref{identification} \quad \Rightarrow \quad \qquad &=\frac{1}{48}\big(\mrd (\star u)\big)_{\mu_{_1}\mu_{_2} \mu_{_3} \mu_{_4}}\big(\mrd (\star u)\big)^{\mu_{_1}\mu_{_2} \mu_{_3} \mu_{_4}}\\
&=\frac{1}{48}\big(\star \nabla_\alpha u^\alpha \big)_{\mu_{_1}\mu_{_2} \mu_{_3} \mu_{_4}}\big(\star \nabla_\beta u^\beta \big)^{\mu_{_1}\mu_{_2} \mu_{_3} \mu_{_4}}\label{proof 1}\\
\eqref{Hodge scalar} \quad \Rightarrow \quad \qquad&=\frac{1}{48}\left(\epsilon_{\mu_{_1}\mu_{_2} \mu_{_3} \mu_{_4}}\nabla_{\alpha}u^{\alpha}\right)\left(\epsilon^{\mu_{_1}\mu_{_2} \mu_{_3} \mu_{_4}}\nabla_{\beta}u^{\beta}\right)\label{proof 2}\\
\eqref{identities} \quad \Rightarrow \quad \qquad &=-\frac{(\nabla_\alpha u^\alpha)^2}{2},
\end{align}
where in the derivation of \eqref{proof 1} we used the coderivative identity
\begin{equation}\label{coderivative}
\mrd (\star u)=-\star (\nabla_\alpha u^\alpha)
\end{equation}
which is correct for an arbitrary 1-form $u$ (see Appendix A in Ref. \cite{Hajian:2015eha} and references therein). The proof of Eq. \eqref{nabla A F} can be achieved following the similar steps:
\begin{align}
F_{\mu{\mu_{_2} \mu_{_3} \mu_{_4}}}F_{\nu}{}^{_{\mu_{_2} \mu_{_3} \mu_{_4}}}&= (\mrd A)_{\mu{\mu_{_2} \mu_{_3} \mu_{_4}}}  \ g_{\nu\sigma} (\mrd A)^{\sigma \mu_{_2} \mu_{_3} \mu_{_4}}\\
\eqref{identification} \quad \Rightarrow \quad \qquad &=\frac{g_{\nu\sigma}}{2}\big( \mrd(\star u)\big)_{\mu{\mu_{_2} \mu_{_3} \mu_{_4}}}\big( \mrd (\star u)\big)^{\sigma \mu_{_2} \mu_{_3} \mu_{_4}}\\
\eqref{coderivative} \quad \Rightarrow \quad \qquad&=\frac{g_{\nu\sigma}}{2}\big(\star \nabla_\alpha u^\alpha \big)_{\mu{\mu_{_2} \mu_{_3} \mu_{_4}}}\big(\star \nabla_\beta u^\beta \big)^{\sigma \mu_{_2} \mu_{_3} \mu_{_4}}\label{proof 4}\\
\eqref{Hodge scalar} \quad \Rightarrow \quad \qquad &=\frac{g_{\nu\sigma}}{2}\left(\epsilon_{\mu_{_2} \mu_{_3} \mu_{_4}\mu}\nabla_{\alpha}u^{\alpha}\right)\left(\epsilon^{\sigma\mu_{_2} \mu_{_3} \mu_{_4}}\nabla_{\beta}u^{\beta}\right)\\
\eqref{identities} \quad \Rightarrow \quad \qquad &=-3(\nabla_\alpha u^\alpha)^2 g_{\mu\nu}\label{F2 value}.
\end{align}
By replacement of the results of the calculations above, we deduce that the field equations of the two theories, and so, their solutions are the same via the Hodge duality. 

Through duality, we have demonstrated the equivalence of the Lagrangians given in Eqs. \eqref{II L} and \eqref{L F2}, and that they both yield the Einstein-$\Lambda$ field equations. However, the on-shell Lagrangians differ from the standard Einstein-$\Lambda$ Lagrangian in \eqref{L Einstein-Lambda} due to a sign discrepancy in the cosmological constant $\Lambda$. This mismatch can be resolved by incorporating a surface term into the action. We now proceed to calculate the aether analogue of this surface term within the shifted Lagrangian \eqref{L F2 shifted}, specifically
\begin{align}
 \nabla_\mu \left(A_{\mu_{_2} \mu_{_3} \mu_{_4}} F^{\mu{\mu_{_2} \mu_{_3} \mu_{_4}}}\right)&=\nabla_\mu \left( A_{\mu_{_2} \mu_{_3} \mu_{_4}} (\mrd A)^{\mu{\mu_{_2} \mu_{_3} \mu_{_4}}})\right)\\
\eqref{identification} \quad \Rightarrow \quad \qquad &=\frac{1}{2}\nabla_\mu \Big( \epsilon_{\mu_{_2} \mu_{_3} \mu_{_4}\alpha}u^\alpha \  \big(\mrd (\star u)\big)^{\mu{\mu_{_2} \mu_{_3} \mu_{_4}}}\Big)\\
\eqref{coderivative} \quad \Rightarrow \quad \qquad &=\frac{-1}{2}\nabla_\mu \Big( \epsilon_{\mu_{_2} \mu_{_3} \mu_{_4}\alpha}u^\alpha \  \big(\star \nabla_\beta u^\beta\big)^{\mu{\mu_{_2} \mu_{_3} \mu_{_4}}}\Big)\\
\eqref{Hodge scalar} \quad \Rightarrow \quad \qquad &=\frac{-1}{2}\nabla_\mu \left( \epsilon_{\mu_{_2} \mu_{_3} \mu_{_4}\alpha}u^\alpha \    \epsilon^{\mu\mu_{_2} \mu_{_3} \mu_{_4}}  \nabla_{\beta}{u^\beta}\right)\\
\eqref{identities} \quad \Rightarrow \quad \qquad &=-3\nabla_\mu \left(u^\mu \nabla_\beta u^\beta\right).
\end{align}
Therefore, the Lagrangian of the minimal Einstein-aether theory can be modified to 
\begin{equation}
\mathcal{L}=R+\mathcal{L}_M-c_2(\nabla_\alpha u^\alpha)^2+2{c_2}\nabla_\mu \left(u^\mu \nabla_\alpha u^\alpha\right)
\end{equation}
in order to reproduce the Einstein-$\Lambda$ Lagrangian in \eqref{L Einstein-Lambda} on-shell.

\section{Some applications of the duality}\label{implications}
The identification of the gauge formulation and the minimal Einstein-aether theory may have different conceptual and practical implications. Here, we mention briefly two of them, and leave more investigations to the later works. 
 
\subsection{Exchanging the solutions}
One usage of the introduced duality is transferring the known solutions from one theory to the other one. In this regard, we note that solutions to the gauge field formulation are more studied because the subject is known from the  1980's. A non-exclusive list of solutions to this theory has been collected in Ref. \cite{Hajian:2021hje}. Here we exemplify one of them to show the simplicity of the procedure. The targeted model here is the Einstein-Maxwell-$\Lambda$ theory in four dimensions \cite{Kerr:1963ud,Newman:1965tw,Newman:1965my,Carter:1968ks,Carter:1970ea,Carter:1970ea2}
\begin{equation}
\mathcal{L}=\frac{1}{16\pi G}(R-2\Lambda-\tilde F_{\mu\nu}\tilde F^{\mu\nu}),
\end{equation}
in which $\tilde F$ is the Maxwell field strength. The solution in our focus is the Kerr-Newman-(Anti) de Sitter black hole. The metric has three integration constants $\{m,a,q\}$, and in Boyer–Lindquist coordinates $x^\mu=(t,r,\theta,\varphi)$ it is
\begin{align}\label{KN AdS}
\mathrm{d}s^2= -&\Delta_\theta(\frac{1-\frac{\Lambda r^2}{3}}{\Xi}-\Delta_\theta f)\mathrm{d}t^2+\frac{\rho ^2}{\Delta_r}\mathrm{d}r^2+\frac{\rho ^2}{\Delta_\theta} \mathrm{d}\theta ^2 -2\Delta_\theta fa\sin ^2 \theta\,\mathrm{d}t \mathrm{d}\varphi\nonumber \\
+&\left( \frac{r^2+a^2}{\Xi}+fa^2\sin ^2\theta \right)\sin ^2\theta\,\mathrm{d}\varphi ^2\,,
\end{align}
where
\begin{align}\label{Rho}
\rho^2 &\equiv r^2+a^2 \cos^2 \theta\,,\qquad \Delta_r \equiv (r^2+a^2)(1-\frac{\Lambda r^2}{3})-2Gmr + q^2\,,\nonumber\\
\Delta_\theta&\equiv 1+\frac{\Lambda a^2}{3}\cos ^2\theta\,,\qquad \Xi\equiv 1+\frac{\Lambda a^2}{3}\,,\qquad
f\equiv\frac{2Gmr-q^2}{\rho ^2\Xi^2}\,.
\end{align}
In these coordinates, the Maxwell gauge field is
\begin{equation}\label{KN AdS A}
\tilde A=\frac{qr}{\rho^2\Xi}(\Delta_\theta\mrd t-a\sin^2 \theta \,\mrd \varphi)\,. 
\end{equation} 
For positive and negative signs of $\Lambda$, the solution is a de Sitter or anti-de Sitter Kerr-Newman black hole, respectively.  Notice that $\Lambda$ in this setup is a coupling constant in contrast with a an integration constant playing the role of a conserved charge. 

The $\Lambda$ can be promoted to be an integration constant and a conserved charge via the gauge formulation that was reviewed before. To this end, the theory is described by the Lagrangian
\begin{equation}
\mathcal{L}=\frac{1}{16\pi G}(R-\tilde F_{\mu\nu}\tilde F^{\mu\nu}\mp 2 F^2),
\end{equation}
on the same footing as Eq. \eqref{L F2}. Equations of motion for this theory are
\begin{equation}\begin{split}
&G_{\mu\nu}={8\pi G}\ T_{\mu\nu}\pm \frac{1}{3} \Big(F_{\mu{\mu_{_2} \mu_{_3} \mu_{_4}}}F_{\nu}{}^{_{\mu_{_2} \mu_{_3} \mu_{_4}}}-3 F^2 g_{\mu\nu}\Big)\\ 
&\nabla_\mu \tilde F^{\mu\nu}=0\\
&\nabla_\mu F^{\mu{\mu_{_2} \mu_{_3} \mu_{_4}}}=0,
\end{split}
\end{equation}
where
\begin{equation}
T_{\mu \nu}=\frac{1}{4\pi G}(\tilde F^{\,\,\alpha}_{\mu}\tilde F_{\nu\alpha}-\frac{1}{4}\tilde F_{\alpha \beta}\tilde F^{\alpha \beta}).
\end{equation}
It can be checked that the same metric and Maxwell gauge field as in Eqs. \eqref{KN AdS} and \eqref{KN AdS A} concomitant with the gauge field
\begin{equation}\label{A Kerr-Newman}
{A}=-\frac{\sqrt{|\Lambda|}(r^3+3ra^2\cos^2\theta+\frac{ma^2}{\Xi})\sin\theta}{3\Xi}\mrd t \wedge\mrd \theta \wedge\mrd \varphi.    
\end{equation}
solve the equations of motion. The gauge freedom has been fixed by demanding that the mass, angular momentum, and other charges be reproduced correctly by the covariant formulation of charges (see Refs. \cite{Hajian:2021hje,Hajian:2016kxx} for more details). In this construction, $\Lambda$ is an integration constant that is in correspondence with the gauge symmetry conserved charge. 

Having the $A$ in hand, it is easy to find the aether field $u^\mu$ by the Hodge duality relation Eq. \eqref{identification}. The result is
\begin{equation}\label{KN AdS u}
u=-\sqrt{2}(\star A) \qquad \Rightarrow \qquad u^\mu= \left(0,\frac{\sqrt{2|\Lambda|}(r^3+3ra^2\cos^2\theta+\frac{ma^2}{\Xi})\sin\theta}{3\Xi \sqrt{-g}},0,0\right). 
\end{equation}
Therefore, the minimal Einstein-aether formulation of the (A)dS-Kerr-Newman example can be explained simply by the Lagrangian
\begin{equation}
\mathcal{L}=R-\tilde F_{\mu\nu}\tilde F^{\mu\nu}\pm(\nabla_\alpha u^\alpha)^2,
\end{equation}
and the field equations
\begin{align}
&G_{\mu\nu}=8\pi G\ T_{\mu\nu}\mp\frac{1}{2}(\nabla_\alpha u^\alpha)^2 g_{\mu\nu},\\
&\nabla_\mu (\nabla_\alpha u^\alpha)=0,
\end{align}
which has the (A)dS-Kerr-Newman metric \eqref{KN AdS} and electromagnetic field \eqref{KN AdS A} besides the aether field \eqref{KN AdS u} as a solution to the field equations. Note that the $\Lambda$ appears as a constant of integration in this setup via
\begin{equation}
{\Lambda}=\pm\frac{1}{2}(\nabla_\alpha u^\alpha)^2.
\end{equation}

It is an interesting line of research to study the solutions of each theory from the point of view of the other theory. However, we do not discuss this issue further, and an interested reader is encouraged to look at the zoo of solutions e.g., in Ref. \cite{Hajian:2021hje}.

\subsection{Discarding pure gauge solutions}
The second application of the aether-gauge duality is to show that some of the solutions that have been proposed to the minimal Einstein-aether theory are pure gauges and hence, they are trivial. To explain the procedure, we focus on the Kerr black hole and the aether field in this geometry that has been introduced in Ref. \cite{Franzin:2023rdl}. The solution is described by the metric and aether fields
\begin{align}
&\mrd s^2= -(1-\frac{2Gmr}{\rho^2})\mrd t^2 -\frac{4Gmra\sin^2\theta}{\rho^2}\mrd t \mrd \varphi +\frac{\rho^2}{\Delta}\mrd r^2 + \rho^2 \mrd \theta^2 +\frac{\Sigma \sin^2\theta}{\rho^2}\mrd \varphi^2,\\
&u_\mu=\left(\sqrt{\frac{\rho^2\Delta+G^4m^4\Theta^2}{\Sigma}},-\frac{G^2m^2\Theta}{\Delta},0,0 \right), \label{example u}
\end{align}   
in which 
\begin{align}
\rho^2\equiv r^2+a^2\cos^2\theta, \qquad \Delta\equiv r^2+a^2-2Gmr, \qquad \Sigma\equiv (r^2+a^2)^2-\Delta a^2\sin^2\theta.
\end{align}
The $\Theta (\theta)$ is an arbitrary function of the coordinate $\theta$. The fields  satisfy the equations of motion \eqref{II EoM g} and \eqref{II EoM u} with $T_{\mu\nu}=0$ and for an arbitrary $c_2$ while the aether field satisfy $\nabla_\alpha u^\alpha=0$. 
By the duality relation \eqref{identification}, we can find the dual gauge field from \eqref{example u} as
{\small
\begin{align}
A=\frac{\sqrt{\frac{\rho^2\Delta+G^4m^4\Theta^2}{2\Sigma}}\sin\theta}{\Delta}\left(2Gmra \ \mrd t\! \wedge\! \mrd r\!\wedge\!\mrd\theta-{\Sigma} \ \mrd r\!\wedge\!\mrd\theta\wedge\! \mrd\varphi\right)+\frac{G^2m^2\Theta\sin\theta}{\sqrt{2}}\  \mrd t\!\wedge\!\mrd\theta\!\wedge\! \mrd\varphi.
\end{align}
}
However, one can check that $\mrd A=0$, which by the Poincar\'e lemma admits $A=\mrd \alpha$ for a (not necessarily unique) 2-form field $\alpha_{\mu\nu}$, and so, $A$ is a pure gauge field. As an example, one can choose 
{\small
\begin{equation}
\alpha=\!\int\! \mrd \theta \left(\frac{\sqrt{\frac{\rho^2\Delta+G^4m^4\Theta^2}{2\Sigma}}\sin\theta}{\Delta}\left(2Gmra \ \mrd t\! \wedge\! \mrd r+\!{\Sigma} \ \mrd r\!\wedge\! \mrd\varphi\right)+\! \frac{G^2m^2\Theta\sin\theta}{\sqrt{2}}\  \mrd t\!\wedge\! \mrd\varphi \right).
\end{equation} 
}
As a result, the field strength vanishes, $F=\mrd A=0$. In this regard, this solution is not physically different from the Kerr metric without any aether field.

Although we showed that the aether field can be removed by a pure gauge, this result is always the case for the special case of $\nabla_\alpha u^\alpha=0$ which has been singled out under the name of ``minimal aether-Einstein theory" in Ref. \cite{Franzin:2023rdl}. 

\noindent \textbf{Theorem:}
\emph{In an aether theory where the parameters $c_1$, $c_3$, and $c_4$ are zero, a divergence-free aether field is necessarily a pure gauge and therefore non-physical.}

\noindent\emph{Proof:} When $c_1 = c_3 = c_4 = 0$, the condition for a divergence-free aether field, $\nabla_\alpha u^\alpha = 0$, is equivalent to $\Lambda = 0$ according to the relation \eqref{tilde Lambda}. In the dual description, using equations \eqref{F on-shell} and \eqref{Lambda c}, $\Lambda = 0$ implies that the field strength $F$ is also zero. Since $F = \mrd A = 0$, the Poincar\'e lemma guarantees the existence of a 2-form field $\alpha_{\mu\nu}$ such that the gauge potential $A$ can be expressed as $A = \mrd \alpha$, indicating a pure gauge configuration.

\section{Conclusion}\label{conclusion}
In this paper we explored a connection between minimal Einstein-aether theory, a gravity theory with a preferred direction, and the gauge field formulation of the cosmological constant. We demonstrated that the aether vector field in its minimal version is Hodge dual to a gauge field, revealing an equivalence between the two theoretical frameworks. This duality allows for the straightforward transfer of known solutions between the theories and provides a mechanism to identify and discard trivial ``pure gauge" solutions within minimal Einstein-aether theory, specifically proving that divergence-less aether fields are non-physical. This latter was exemplified in the Kerr black hole background and was proved as a theorem in general.  The identified equivalence offers new perspectives on both theories, including potential implications for cosmological solutions and black hole physics.

\noindent \textbf{Acknowledgements:} The author gratefully acknowledges the kind support of Jutta Kunz at Oldenburg University and Bayram Tekin at Middle East Technical University. Furthermore, the author wishes to thank the organizers of the ``3rd online international conference; Frontiers in General Relativity, Modified Theories, and Cosmology" at the University of Tabriz, as well as Prof. Çetin Şentürk for an insightful presentation on the minimal aether theory, which served as the initial inspiration for this research. This work was supported by the TÜBİTAK International Researchers Program under Grant No. 2221.

\end{document}